\documentclass[aip,apl,amsmath,amssymb,reprint,superscriptaddress]{revtex4-1}
\usepackage{graphicx}
\usepackage{amsmath}
\usepackage{dcolumn}
\usepackage{bm}
\usepackage{bbold}
\usepackage{tabularx}

\bibstyle{apsrev4-1}

\begin{document}

\title{A 3D Printed Superconducting Aluminium Microwave Cavity}

\author{Daniel L. Creedon}
\affiliation{School of Physics, University of Melbourne, Parkville, Victoria 3010, Australia}

\author{Maxim Goryachev}
\affiliation{ARC Centre of Excellence for Engineered Quantum Systems, University of Western Australia, 35 Stirling Highway, Crawley WA 6009, Australia}

\author{Nikita Kostylev}
\affiliation{ARC Centre of Excellence for Engineered Quantum Systems, University of Western Australia, 35 Stirling Highway, Crawley WA 6009, Australia}

\author{Tim Sercombe}
\affiliation{School of Mechanical and Chemical Engineering, University of Western Australia, 35 Stirling Hwy, Crawley 6009, Australia}

\author{Michael E. Tobar}
\email{michael.tobar@uwa.edu.au}
\affiliation{ARC Centre of Excellence for Engineered Quantum Systems, University of Western Australia, 35 Stirling Highway, Crawley WA 6009, Australia}

\date{\today}


\begin{abstract}
3D printing of plastics, ceramics, and metals has existed for several decades and has revolutionized many areas of manufacturing and science. Printing of metals in particular has found a number of applications in fields as diverse as customized medical implants, jet engine bearings, and rapid prototyping in the automotive industry. Whilst many techniques can be used for 3D printing metals, they commonly rely on computer controlled melting or sintering of a metal alloy powder using a laser or electron beam. The mechanical properties of parts produced in such a way have been well studied, but little attention has been paid to their electrical properties. Here we show that a microwave cavity (resonant frequencies 9.9 and 11.2 GHz) 3D printed using an Al-12Si alloy exhibits superconductivity when cooled below the critical temperature of aluminium (1.2 K), with a performance comparable to the common 6061 alloy of aluminium. Superconducting cavities find application in numerous areas of physics, from particle accelerators to cavity quantum electrodynamics experiments.  The result is achieved even with a very large concentration of non-superconducting silicon in the alloy of $12.18\%$, compared to Al-6061, which has between $0.4$ to $0.8\%$. Our results may pave the way for the possibility of 3D printing superconducting cavity configurations that are otherwise impossible to machine.\end{abstract}

\maketitle

Superconducting cavities are commonly used to trap and store resonant microwave radiation and reduce losses, allowing devices with very high quality factor ($Q$), narrow bandwidth and long storage times\cite{turneareJAP68}. Such cavities are essential for many physics applications, including particle accelerators\cite{Grimm05}, ultra-sensitive motion/displacement sensing\cite{Blair95}, precise frequency stabilisation\cite{tw} and testing fundamental physics. In particular tests of the speed of light and the constancy of fundamental constants\cite{tw,Lipa,nagel15} as well as the search for hidden sector particles, which are dark matter candidates\cite{HSP2013a,HSP2013,HSP2011,jjar2010} depend on such cavities. More recently they have been applied to cavity quantum electrodynamics (CQED) experiments to house qubit devices and provide a reduced density of states to radiate into, vital for the success of such experiments. The use of these cavities allows enhanced coherence times\cite{PaikPRL} and may even serve as a quantum memory. Superconducting cavities are often precision-machined from extremely high purity aluminium or niobium at great cost and investment of time in achieving optimal surface preparation.  

The use of highly pure aluminium is not always necessary,  as the standard 6061 aluminium alloy allows very high mechanical Q-factors, and was used in the past for large scale acoustic resonant-mass gravitational wave detectors\cite{lsu}. The alloy has been shown to become superconducting, allowing microwave quality factors on the order of 10$^6$ to be achieved\cite{reagor}. An analysis by Reagor et al.\cite{reagor} shows that limitations in quality factor are dependent on dielectric losses from the oxide layer lining the interior of the cavity, as well as damping from finite residual surface resistance, both effects being dependent on the spatial distribution of the choice of mode excited in the cavity. 

With the rise in popularity of 3D printing in recent years, it is now fast and inexpensive to manufacture parts from metallic powders, commonly an aluminium alloy with a high concentration of silicon to allow effective laser melting. To investigate whether 3D printed aluminium-silicon cavities exhibited superconductivity, a simple resonant microwave cavity was designed in a cylindrical configuration. The cavity was modelled using COMSOL Multiphysics finite element analysis software, and the geometric factor for the modes computed using:

\begin{equation}
G = \frac{\omega \mu_0 \iiint \lvert \overrightarrow{H} \rvert^2 dV}{\iint \lvert \overrightarrow{H} \rvert^2 dS}
\end{equation}

where $\omega$ is the (angular) eigenfrequency of the mode, $\mu_0$ is the vacuum permeability, and $\overrightarrow{H}$ is the magnetic field component of the mode. The geometric factor is a cavity parameter which is independent of the cavity size and wall losses, and is related to the intrinsic quality factor of the cavity through the surface resistance $R_S$:

\begin{equation}
Q_0 = \frac{G}{R_S}
\end{equation}

\begin{figure}[!t]
\begin{center}
\includegraphics[width=\columnwidth]{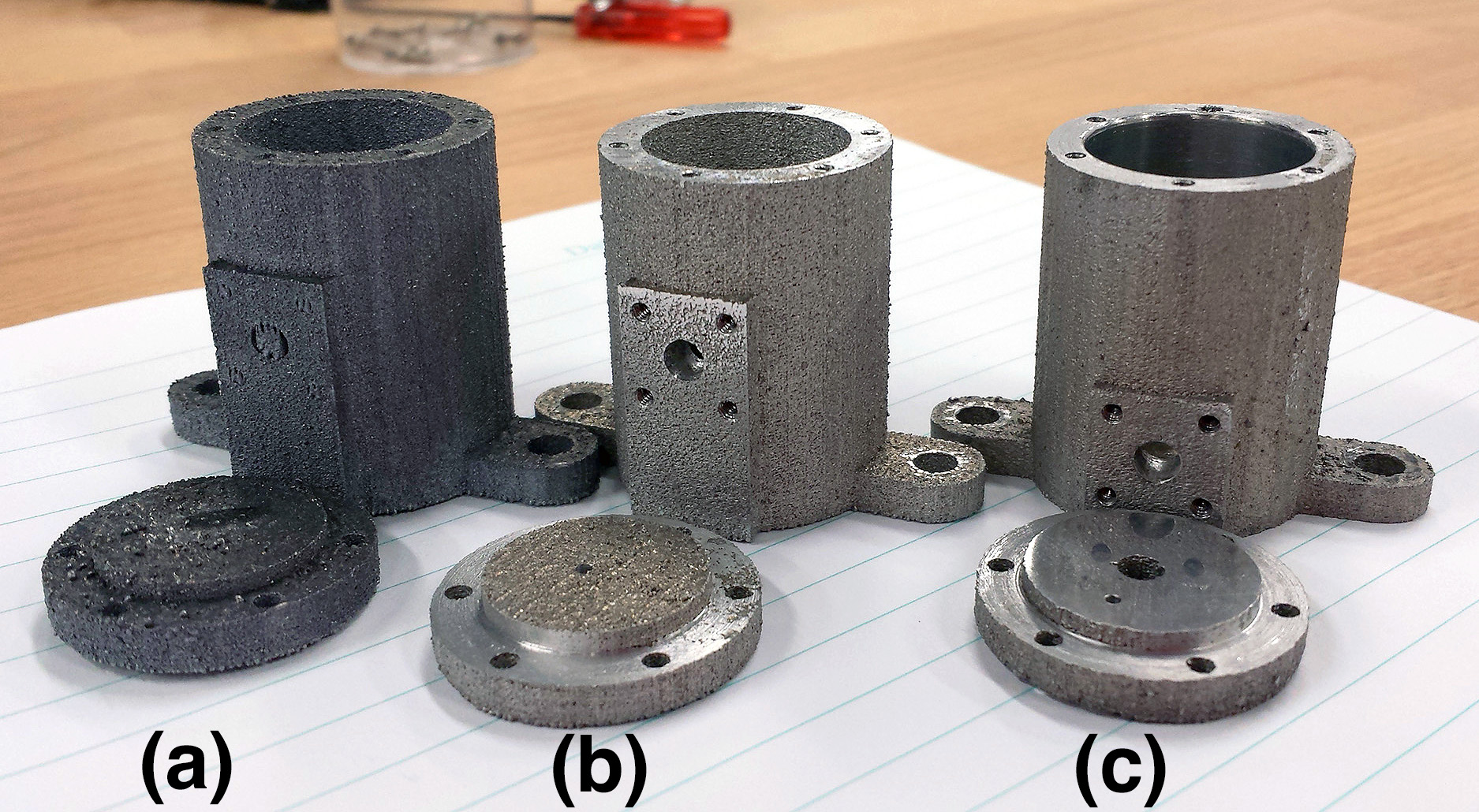}
\caption{Photograph of several samples of the 3D printed cavity. The colour difference between the non-annealed and annealed samples is readily visible. (a) Cavity as-printed, (b) Cavity after annealing and having microwave probe holes drilled and tapped and mating surfaces of lid and body machined flat, (c) Cavity after annealing, machining, and polishing.}
\label{cavitydesigns}
\end{center}
\end{figure}

Two versions of the cavity were created, differing only in the location of the probe mounts, as seen in Fig. \ref{cavitydesigns}, with the internal dimensions of 20mm diameter and 30mm height remaining nominally the same.   The cavities were 3D printed using a Selective Laser Melting process, performed on a Realizer SLM 100 (ReaLizer GmbH, Germany), which was equipped with a fibre laser ($\lambda$=1.06$\mu$m) having a maximum power of 200 W at the part bed. Al-12Si (in wt.\%) powder ($d_{50}\sim38\mu$m, TLS Technik, Germany) was used. The exact composition of the powder is 12.18\% silicon, 0.118\% iron, 0.003\% copper, with the balance consisting of aluminium. In contrast, 6061 aluminium alloy typically contains maximum impurity composition of 0.8\% silicon, 0.7\% iron, 0.15\% copper and 1.2\% magnesium. An inert, high purity argon gas atmosphere was used during processing to minimise oxidation. The laser scan speed and laser power were 1000 mm/s and 200 W, respectively. The powder layer thickness was fixed at 50 $\mu$m and the scan spacing at 150 $\mu$m, while the substrate was heated to 200$^{\circ}$C.

\begin{figure}[b]
\begin{center}
\includegraphics[width=3.4in]{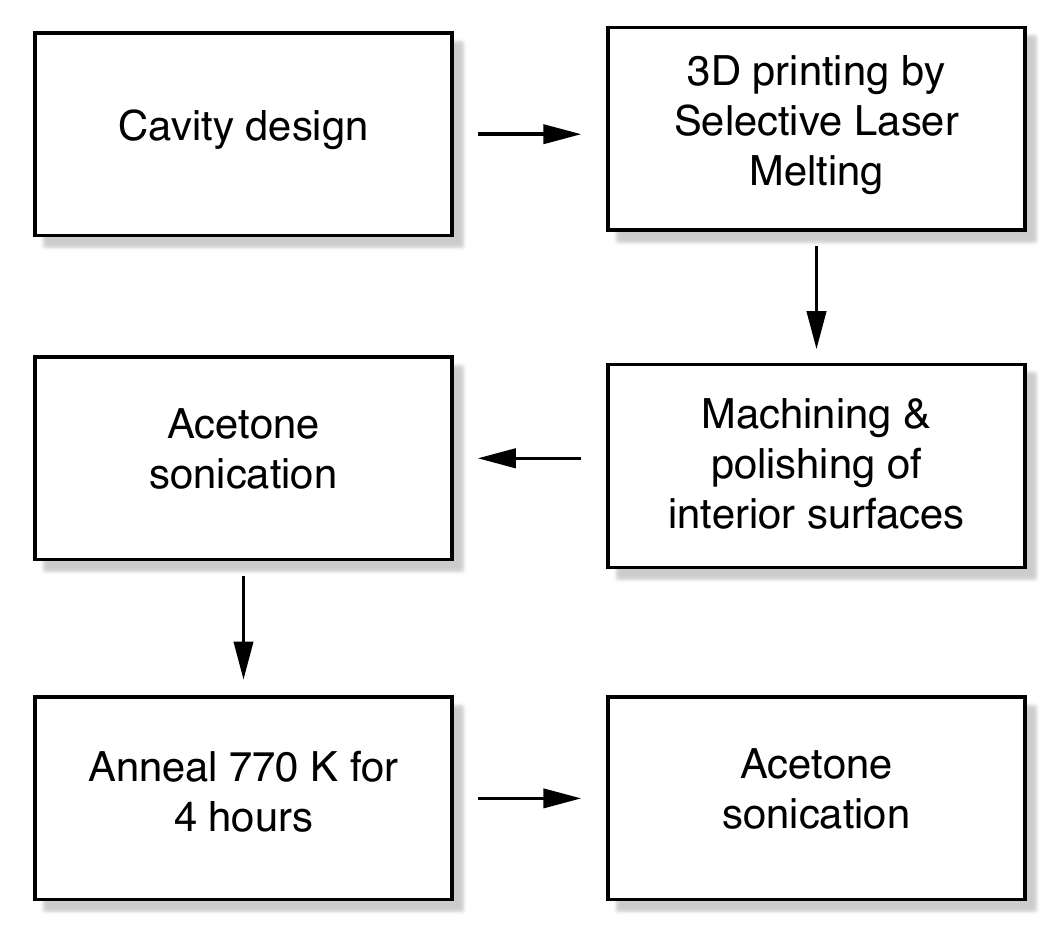}
\caption{Simple flow chart showing an example of the fabrication steps for the cavities described.}
\label{flowch}
\end{center}
\end{figure}

After printing, the cavities underwent further processing.  In addition to being measured bare (as-printed), the internal surfaces of the cavity were also machined smooth, then wet sanded with 220, 600, and 1000 grit sandpaper, followed by polishing with a felt-tipped rotary tool and diamond paste at 1200, 1800, 3000, 14000, and 50000 grit. The cavity was wiped clean with acetone in between steps, and finally submerged in acetone in an ultrasonic bath for 1 hour and thoroughly cleaned. A schematic flow chart showing the fabrication process is given in Fig. \ref{flowch}. Microwave energy was coupled in and out of the cavity using small `loop' probes consisting of coaxial microwave cable with the outer conductor and insulation stripped, and the center conductor formed back in a loop and soldered to the outer conductor.

The cavities were bolted to an OFHC (oxygen free high thermal conductivity) copper rod attached to the mixing chamber of a cryogen-free dilution refrigerator. This system is capable of achieving a base temperature of less than 20 millikelvin, and is necessary in order to cool the samples below the known superconducting temperature of 1.2 K for aluminium. The surface resistance of the cavity walls was then determined by converting, using the geometric factor, from a direct measurement of the cavity quality factor made with a vector network analyser (VNA). The VNA was connected to the cryostat using coaxial microwave cables which were thermally anchored at each stage of the dilution refrigerator, and the cavity then measured in transmission, with the weak outcoming signal amplified by a low-noise cryogenic microwave amplifier situated at the 4K stage of the system. Measurements were first performed with the bare cavities (as printed), followed by machining and polishing smooth the internal surfaces of the cavities.

Coupling to the cavities such that the superconducting transition could be measured posed a number of challenges.  Because the ratio of quality factor to coupling remains constant in such cavities, the probes must be prepared in such a way that the mode of interest is severely undercoupled to the point of being unmeasurable at room temperature. In this way, the many orders of magnitude increase in coupling associated with the superconducting transition still results in a very low-coupled mode where the quality factor is as close to intrinsic as possible, i.e. not loaded by the coupling probes. This also ensures that a reduced vibration sensitivity is achieved, as small displacements of probes with coupling set close to unity can result in very large change in coupling and thus detected amplitude. An insertion loss of -80 to -90 dB at the peak of the mode at room temperature was found to give good results. To achieve such low coupling, the loop probes were inserted in the wrong orientation intentionally - i.e. with the plane of the loop parallel to the magnetic field vector rather than perpendicular, resulting in an extremely small coupling to the mode. 

\begin{figure}[htbp]
\begin{center}
\includegraphics[width=3.0in]{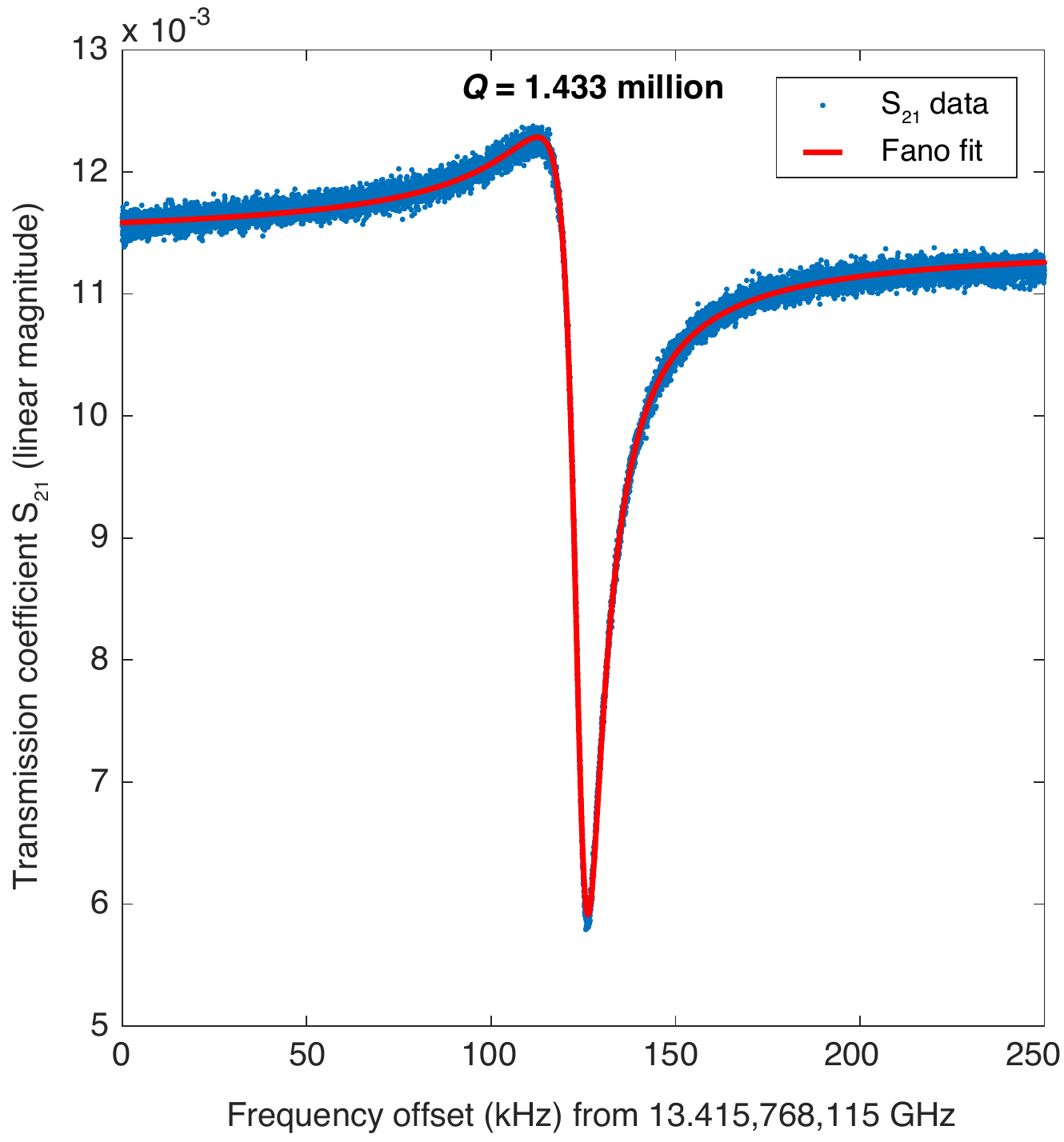}
\caption{An example of a Fano resonance fit to the transmission coefficient of the TE$_{1,1,2}$ mode below 100 mK in temperature, which is used to extract the resonance frequency and Q-factor.}
\label{high-Q}
\end{center}
\end{figure}

\begin{figure}[htbp]
\begin{center}
\includegraphics[width=3.4in]{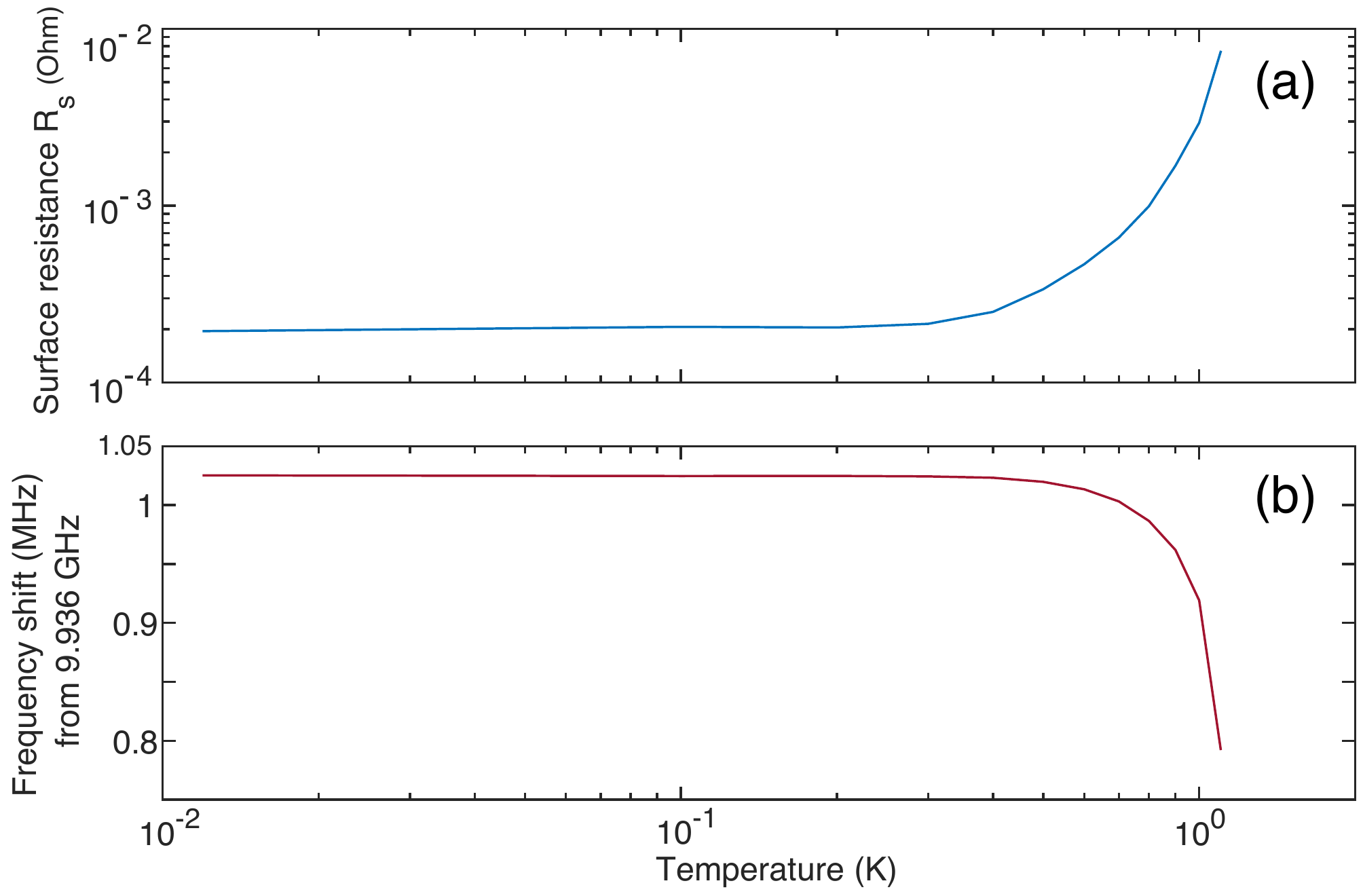}
\caption{(a) Surface resistance of a 3D printed cavity with polished internal surfaces and extremely low coupling as measured through the TE$_{1,1,1}$ mode at 9.9 GHz, and (b) frequency shift of the same mode as a function of temperature. Because the coupling must necessarily be set extremely low to allow an intrinsic measure of quality factor, the mode cannot be observed above the superconducting transition temperature of 1.2 K.}
\label{rsvstemp}
\end{center}
\end{figure}

In the cavity configuration selected, a number of modes were found that could be coupled to.  Modelling the twelve lowest order modes in COMSOL Multiphysics (see Table \ref{tableofmode}) showed a large variation in geometric factor of the modes, indicating that cavity Q-factor is not an accurate comparison between different modes. Thus, we compute the surface resistance $R_S$ using the geometric factor and measured quality factor.  Based on the choice of probe location and the aforementioned difficulties in setting the coupling to an appropriately small value at room temperature, the modes TE$_{1,1,1}$ at 9.9 GHz, TE$_{0,1,0}$ at 11.2 GHz and TE$_{1,1,2}$ at 13.4 GHz were selected, as these were found to be the easiest to couple to.  Somewhat surprisingly, it was found that leaving the cavity walls rough, i.e. as-printed by the SLM machine, had no disadvantage compared to machining the internal surfaces of the cavity smooth and polishing with diamond paste. Any difference observed resulted in an improvement of $\sim$10\%, which is within the error of the Q-factor measurement.

Following these measurements, annealing of the cavity at 770 K for 4 hours was performed.  This was found to have a larger effect, resulting in an improvement in quality factor for the highly polished cavity, but having no effect for the rough `as-printed' cavity. We anticipate that the losses here are dominated by the surface roughness and thus no improvement in Q-factor is seen after annealing. In the best case, the Q-factor of a mode was improved from 1.6 million to 3.8 million. After 3D printing, the silicon content of the cavity is approximately 6-7\%, however the solubility of Si in Al is only $\sim$1.5\%, and nearly zero at room temperature.  Annealing at high temperature followed by slowly letting the cavity cool to room temperature was found to drive off most of the silicon, leaving a pure aluminium matrix behind.  The brightness of the cavity finish was also noticeably changed from dark to light grey after annealing.

Fig. \ref{high-Q} shows the $S_{21}$ transmission coefficient of the TE$_{1,1,2}$ mode, which is well modelled by a Fano resonance fit\cite{RevModPhys.82.2257,PhysRevB.82.195307} - a pure Lorentzian lineshape profile with an asymmetry factor. The Q-factor of each mode was computed by fitting a Fano resonance to the transmission curve, and taking the ratio of center frequency and linewidth extracted from the Lorentzian part of the fit.

Fig. \ref{rsvstemp} shows the surface resistance (as calculated from the measured Q-factor and modelled geometric factor) and frequency shift of the TE$_{1,1,1}$ mode as a function of temperature for an internally polished cavity. At 20 millikelvin temperature a Q-factor of $1.5\times10^6$ was obtained.

\begin{table}[]
\centering
\label{tableofmode}
\begin{tabularx}{\columnwidth}{|X|X|X|}
\hline
Frequency (GHz) & Mode & Geometric Factor \\
\hline
\hline
10.0	&	TE$_{\text{(1,1,1)}}$	&	300.4        \\
11.2	&	TM$_{\text{(0,1,0)}}$	&	334.5         \\
12.35	&	TM$_{\text{(0,1,1)}}$	&	292        	\\
13.4	&	TE$_{\text{(1,1,2)}}$	&	432.5         \\
15.2	&	TE$_{\text{(2,1,1)}}$	&	353.8        \\
15.2	&	TM$_{\text{(0,1,2)}}$	&	358.5        \\
17.6	&	TE$_{\text{(2,1,2)}}$	&	426.5        \\
17.6	&	TE$_{\text{(1,1,3)}}$	&	592.4        \\
17.9	&	TM$_{\text{(1,1,0)}}$	&	528.9         \\
18.6	&	TE$_{\text{(0,1,1)}}$	&	755.9        \\
19.0	&	TM$_{\text{(0,1,3)}}$	&	447           \\
20.6	&	TE$_{\text{(0,1,2)}}$	&	885.3       \\
\hline
\end{tabularx}
\caption{Summary of lowest order modes and their geometric factors as modelled using COMSOL Multiphysics}
\end{table}

In order to determine the superconducting critical magnetic field of the cavities, an external DC magnetic field was applied parallel to the cavity axial direction. The magnetic sensitivity of the cavities was found to be in good agreement with the known critical field for aluminium, with the superconducting effect being destroyed and the quality factor of the mode degraded to the point of the mode disappearing entirely after the application of a field greater than 10 mT . Even after ramping the field back to zero, the mode could not be recovered until the cavity was heated above its superconducting critical temperature of 1.2K to release trapped magnetic flux.  Modes that were highly overcoupled remained visible with the application of magnetic field, and underwent no change in quality factor.  For instance, when highly overcoupled, we found that the TE$_{1,1,1}$ mode in an as-printed cavity exhibits Q$\sim1.5\times 10^4$, a value that remained constant up to applied external DC magnetic field values of in excess of 5 T.  This may be important for axion detection experiments, which require very large cavities in high magnetic field with quality factors on the order of the axion linewidth of 10$^6$.  The mode measured here corresponds to Q$\sim 7\times 10^4$ at the axion search frequency of 500 MHz, given the constant ratio of quality factor to frequency.\\

In summary, we show that 3D printing of an aluminium microwave cavity using commonly sourced `dirty' aluminium powder allows the observation of superconductivity in the resultant metal. The results are comparable to cavities machined in the conventional way from common Al-6061 alloy\cite{reagor}, and are unaffected by the surface roughness of the cavity walls due to the 3D printing process.  A combination of machining and polishing smooth the cavity interior, followed by annealing at 770 K for 4 hours to drive off residual silicon impurities was found to improve the Q-factor by approximately a factor of two.  We anticipate that the use of highly pure metallic powders, rather than the Al-12Si powder used in the present work, will result in further improvements comparable to state of the art aluminium cavities manufactured in the conventional way. Such cavities have been shown to reach quality factors on the order of 0.5 $\times 10^9$ for similar mode frequencies when highly pure 5N5 aluminium is used\cite{reagor}.  Indeed, we note that recently published results\cite{niobium} have shown that using highly pure Niobium powder and an a different 3D printing process known as Electron Beam Melting, when combined with advanced surface preparation techniques such as buffered chemical polishing, can result in quality factors on the order of $10^9$ for cavity frequencies of 3.8 GHz.  For applications where such high quality factors are not required, the process of SLM with the relatively much more inexpensive Al-12Si powder may prove a useful and rapid way to fabricate experimental apparatus.  Little experimental data exists in the literature for 3D printed superconducting cavities of any description, and thus further work will be required to determine the optimum material and technique for additive manufacturing of such devices.

\begin{acknowledgments}
This work was supported by the Australian Research Council grant number CE110001013.
\end{acknowledgments}
\bibliography{biblio}

\begin{thebibliography}{16}%
\makeatletter
\providecommand \@ifxundefined [1]{%
 \@ifx{#1\undefined}
}%
\providecommand \@ifnum [1]{%
 \ifnum #1\expandafter \@firstoftwo
 \else \expandafter \@secondoftwo
 \fi
}%
\providecommand \@ifx [1]{%
 \ifx #1\expandafter \@firstoftwo
 \else \expandafter \@secondoftwo
 \fi
}%
\providecommand \natexlab [1]{#1}%
\providecommand \enquote  [1]{``#1''}%
\providecommand \bibnamefont  [1]{#1}%
\providecommand \bibfnamefont [1]{#1}%
\providecommand \citenamefont [1]{#1}%
\providecommand \href@noop [0]{\@secondoftwo}%
\providecommand \href [0]{\begingroup \@sanitize@url \@href}%
\providecommand \@href[1]{\@@startlink{#1}\@@href}%
\providecommand \@@href[1]{\endgroup#1\@@endlink}%
\providecommand \@sanitize@url [0]{\catcode `\\12\catcode `\$12\catcode
  `\&12\catcode `\#12\catcode `\^12\catcode `\_12\catcode `\%12\relax}%
\providecommand \@@startlink[1]{}%
\providecommand \@@endlink[0]{}%
\providecommand \url  [0]{\begingroup\@sanitize@url \@url }%
\providecommand \@url [1]{\endgroup\@href {#1}{\urlprefix }}%
\providecommand \urlprefix  [0]{URL }%
\providecommand \Eprint [0]{\href }%
\providecommand \doibase [0]{http://dx.doi.org/}%
\providecommand \selectlanguage [0]{\@gobble}%
\providecommand \bibinfo  [0]{\@secondoftwo}%
\providecommand \bibfield  [0]{\@secondoftwo}%
\providecommand \translation [1]{[#1]}%
\providecommand \BibitemOpen [0]{}%
\providecommand \bibitemStop [0]{}%
\providecommand \bibitemNoStop [0]{.\EOS\space}%
\providecommand \EOS [0]{\spacefactor3000\relax}%
\providecommand \BibitemShut  [1]{\csname bibitem#1\endcsname}%
\let\auto@bib@innerbib\@empty
\bibitem [{\citenamefont {Turneaure}\ and\ \citenamefont
  {Weissman}(1968)}]{turneareJAP68}%
  \BibitemOpen
  \bibfield  {author} {\bibinfo {author} {\bibfnamefont {J.~P.}\ \bibnamefont
  {Turneaure}}\ and\ \bibinfo {author} {\bibfnamefont {I.}~\bibnamefont
  {Weissman}},\ }\href@noop {} {\bibfield  {journal} {\bibinfo  {journal}
  {Journal of Applied Physics}\ }\textbf {\bibinfo {volume} {39}},\ \bibinfo
  {pages} {4417} (\bibinfo {year} {1968})}\BibitemShut {NoStop}%
\bibitem [{\citenamefont {Grimm}\ \emph {et~al.}(2005)\citenamefont {Grimm},
  \citenamefont {Aizaz}, \citenamefont {Johnson}, \citenamefont {Hartung},
  \citenamefont {Marti}, \citenamefont {Meidlinger}, \citenamefont
  {Meidlinger}, \citenamefont {Popielarski},\ and\ \citenamefont
  {York}}]{Grimm05}%
  \BibitemOpen
  \bibfield  {author} {\bibinfo {author} {\bibfnamefont {T.~L.}\ \bibnamefont
  {Grimm}}, \bibinfo {author} {\bibfnamefont {A.}~\bibnamefont {Aizaz}},
  \bibinfo {author} {\bibfnamefont {M.}~\bibnamefont {Johnson}}, \bibinfo
  {author} {\bibfnamefont {W.}~\bibnamefont {Hartung}}, \bibinfo {author}
  {\bibfnamefont {F.}~\bibnamefont {Marti}}, \bibinfo {author} {\bibfnamefont
  {D.}~\bibnamefont {Meidlinger}}, \bibinfo {author} {\bibfnamefont
  {M.}~\bibnamefont {Meidlinger}}, \bibinfo {author} {\bibfnamefont
  {J.}~\bibnamefont {Popielarski}}, \ and\ \bibinfo {author} {\bibfnamefont
  {R.~C.}\ \bibnamefont {York}},\ }\href@noop {} {\bibfield  {journal}
  {\bibinfo  {journal} {IEEE Transactions on Applied Superconductivity}\
  }\textbf {\bibinfo {volume} {15}},\ \bibinfo {pages} {2393} (\bibinfo {year}
  {2005})}\BibitemShut {NoStop}%
\bibitem [{\citenamefont {Blair}\ \emph {et~al.}(1995)\citenamefont {Blair},
  \citenamefont {Ivanov}, \citenamefont {Tobar}, \citenamefont {Turner},
  \citenamefont {van Kann},\ and\ \citenamefont {Heng}}]{Blair95}%
  \BibitemOpen
  \bibfield  {author} {\bibinfo {author} {\bibfnamefont {D.~G.}\ \bibnamefont
  {Blair}}, \bibinfo {author} {\bibfnamefont {E.~N.}\ \bibnamefont {Ivanov}},
  \bibinfo {author} {\bibfnamefont {M.~E.}\ \bibnamefont {Tobar}}, \bibinfo
  {author} {\bibfnamefont {P.~J.}\ \bibnamefont {Turner}}, \bibinfo {author}
  {\bibfnamefont {F.}~\bibnamefont {van Kann}}, \ and\ \bibinfo {author}
  {\bibfnamefont {I.~S.}\ \bibnamefont {Heng}},\ }\href@noop {} {\bibfield
  {journal} {\bibinfo  {journal} {Phys. Rev. Lett.}\ }\textbf {\bibinfo
  {volume} {74}},\ \bibinfo {pages} {1908} (\bibinfo {year}
  {1995})}\BibitemShut {NoStop}%
\bibitem [{\citenamefont {Turneaure}\ \emph {et~al.}(1983)\citenamefont
  {Turneaure}, \citenamefont {Will}, \citenamefont {Farrell}, \citenamefont
  {Mattison},\ and\ \citenamefont {Vessot}}]{tw}%
  \BibitemOpen
  \bibfield  {author} {\bibinfo {author} {\bibfnamefont {J.~P.}\ \bibnamefont
  {Turneaure}}, \bibinfo {author} {\bibfnamefont {C.~M.}\ \bibnamefont {Will}},
  \bibinfo {author} {\bibfnamefont {B.~F.}\ \bibnamefont {Farrell}}, \bibinfo
  {author} {\bibfnamefont {E.~M.}\ \bibnamefont {Mattison}}, \ and\ \bibinfo
  {author} {\bibfnamefont {R.~F.~C.}\ \bibnamefont {Vessot}},\ }\href@noop {}
  {\bibfield  {journal} {\bibinfo  {journal} {Phys. Rev. D}\ }\textbf {\bibinfo
  {volume} {27}},\ \bibinfo {pages} {1705} (\bibinfo {year}
  {1983})}\BibitemShut {NoStop}%
\bibitem [{\citenamefont {Lipa}\ \emph {et~al.}(2003)\citenamefont {Lipa},
  \citenamefont {Nissen}, \citenamefont {Wang}, \citenamefont {Stricker},\ and\
  \citenamefont {Avaloff}}]{Lipa}%
  \BibitemOpen
  \bibfield  {author} {\bibinfo {author} {\bibfnamefont {J.~A.}\ \bibnamefont
  {Lipa}}, \bibinfo {author} {\bibfnamefont {J.~A.}\ \bibnamefont {Nissen}},
  \bibinfo {author} {\bibfnamefont {S.}~\bibnamefont {Wang}}, \bibinfo {author}
  {\bibfnamefont {D.~A.}\ \bibnamefont {Stricker}}, \ and\ \bibinfo {author}
  {\bibfnamefont {D.}~\bibnamefont {Avaloff}},\ }\href@noop {} {\bibfield
  {journal} {\bibinfo  {journal} {Phys. Rev. Lett.}\ }\textbf {\bibinfo
  {volume} {90}},\ \bibinfo {pages} {060403} (\bibinfo {year}
  {2003})}\BibitemShut {NoStop}%
\bibitem [{\citenamefont {Nagel}\ \emph {et~al.}(2015)\citenamefont {Nagel},
  \citenamefont {Parker}, \citenamefont {Kovalchuk}, \citenamefont {Stanwix},
  \citenamefont {Hartnett}, \citenamefont {Ivanov}, \citenamefont {Peters},\
  and\ \citenamefont {Tobar}}]{nagel15}%
  \BibitemOpen
  \bibfield  {author} {\bibinfo {author} {\bibfnamefont {M.}~\bibnamefont
  {Nagel}}, \bibinfo {author} {\bibfnamefont {S.~R.}\ \bibnamefont {Parker}},
  \bibinfo {author} {\bibfnamefont {E.~V.}\ \bibnamefont {Kovalchuk}}, \bibinfo
  {author} {\bibfnamefont {P.~L.}\ \bibnamefont {Stanwix}}, \bibinfo {author}
  {\bibfnamefont {J.~G.}\ \bibnamefont {Hartnett}}, \bibinfo {author}
  {\bibfnamefont {E.~N.}\ \bibnamefont {Ivanov}}, \bibinfo {author}
  {\bibfnamefont {A.}~\bibnamefont {Peters}}, \ and\ \bibinfo {author}
  {\bibfnamefont {M.~E.}\ \bibnamefont {Tobar}},\ }\href@noop {} {\bibfield
  {journal} {\bibinfo  {journal} {Nat Commun}\ }\textbf {\bibinfo {volume}
  {6}},\ \bibinfo {pages} {10.1038/ncomms9174} (\bibinfo {year}
  {2015})}\BibitemShut {NoStop}%
\bibitem [{\citenamefont {Parker}\ \emph {et~al.}(2013)\citenamefont {Parker},
  \citenamefont {Hartnett}, \citenamefont {Povey},\ and\ \citenamefont
  {Tobar}}]{HSP2013a}%
  \BibitemOpen
  \bibfield  {author} {\bibinfo {author} {\bibfnamefont {S.~R.}\ \bibnamefont
  {Parker}}, \bibinfo {author} {\bibfnamefont {J.~G.}\ \bibnamefont
  {Hartnett}}, \bibinfo {author} {\bibfnamefont {R.~G.}\ \bibnamefont {Povey}},
  \ and\ \bibinfo {author} {\bibfnamefont {M.~E.}\ \bibnamefont {Tobar}},\
  }\href@noop {} {\bibfield  {journal} {\bibinfo  {journal} {Phys. Rev. D}\
  }\textbf {\bibinfo {volume} {88}},\ \bibinfo {pages} {112004} (\bibinfo
  {year} {2013})}\BibitemShut {NoStop}%
\bibitem [{\citenamefont {Parker}, \citenamefont {Rybka},\ and\ \citenamefont
  {Tobar}(2013)}]{HSP2013}%
  \BibitemOpen
  \bibfield  {author} {\bibinfo {author} {\bibfnamefont {S.~R.}\ \bibnamefont
  {Parker}}, \bibinfo {author} {\bibfnamefont {G.}~\bibnamefont {Rybka}}, \
  and\ \bibinfo {author} {\bibfnamefont {M.~E.}\ \bibnamefont {Tobar}},\
  }\href@noop {} {\bibfield  {journal} {\bibinfo  {journal} {Phys. Rev. D}\
  }\textbf {\bibinfo {volume} {87}},\ \bibinfo {pages} {115008} (\bibinfo
  {year} {2013})}\BibitemShut {NoStop}%
\bibitem [{\citenamefont {Povey}, \citenamefont {Hartnett},\ and\ \citenamefont
  {Tobar}(2011)}]{HSP2011}%
  \BibitemOpen
  \bibfield  {author} {\bibinfo {author} {\bibfnamefont {R.~G.}\ \bibnamefont
  {Povey}}, \bibinfo {author} {\bibfnamefont {J.~G.}\ \bibnamefont {Hartnett}},
  \ and\ \bibinfo {author} {\bibfnamefont {M.~E.}\ \bibnamefont {Tobar}},\
  }\href@noop {} {\bibfield  {journal} {\bibinfo  {journal} {Phys. Rev. D}\
  }\textbf {\bibinfo {volume} {84}},\ \bibinfo {pages} {055023} (\bibinfo
  {year} {2011})}\BibitemShut {NoStop}%
\bibitem [{\citenamefont {Jaeckel}\ and\ \citenamefont
  {Ringwald}(2010)}]{jjar2010}%
  \BibitemOpen
  \bibfield  {author} {\bibinfo {author} {\bibfnamefont {J.}~\bibnamefont
  {Jaeckel}}\ and\ \bibinfo {author} {\bibfnamefont {A.}~\bibnamefont
  {Ringwald}},\ }\href@noop {} {\bibfield  {journal} {\bibinfo  {journal}
  {Annual Review of Nuclear and Particle Science}\ }\textbf {\bibinfo {volume}
  {60}},\ \bibinfo {pages} {405} (\bibinfo {year} {2010})}\BibitemShut
  {NoStop}%
\bibitem [{\citenamefont {Paik}\ \emph {et~al.}(2011)\citenamefont {Paik},
  \citenamefont {Schuster}, \citenamefont {Bishop}, \citenamefont {Kirchmair},
  \citenamefont {Catelani}, \citenamefont {Sears}, \citenamefont {Johnson},
  \citenamefont {Reagor}, \citenamefont {Frunzio}, \citenamefont {Glazman},
  \citenamefont {Girvin}, \citenamefont {Devoret},\ and\ \citenamefont
  {Schoelkopf}}]{PaikPRL}%
  \BibitemOpen
  \bibfield  {author} {\bibinfo {author} {\bibfnamefont {H.}~\bibnamefont
  {Paik}}, \bibinfo {author} {\bibfnamefont {D.~I.}\ \bibnamefont {Schuster}},
  \bibinfo {author} {\bibfnamefont {L.~S.}\ \bibnamefont {Bishop}}, \bibinfo
  {author} {\bibfnamefont {G.}~\bibnamefont {Kirchmair}}, \bibinfo {author}
  {\bibfnamefont {G.}~\bibnamefont {Catelani}}, \bibinfo {author}
  {\bibfnamefont {A.~P.}\ \bibnamefont {Sears}}, \bibinfo {author}
  {\bibfnamefont {B.~R.}\ \bibnamefont {Johnson}}, \bibinfo {author}
  {\bibfnamefont {M.~J.}\ \bibnamefont {Reagor}}, \bibinfo {author}
  {\bibfnamefont {L.}~\bibnamefont {Frunzio}}, \bibinfo {author} {\bibfnamefont
  {L.~I.}\ \bibnamefont {Glazman}}, \bibinfo {author} {\bibfnamefont {S.~M.}\
  \bibnamefont {Girvin}}, \bibinfo {author} {\bibfnamefont {M.~H.}\
  \bibnamefont {Devoret}}, \ and\ \bibinfo {author} {\bibfnamefont {R.~J.}\
  \bibnamefont {Schoelkopf}},\ }\href@noop {} {\bibfield  {journal} {\bibinfo
  {journal} {Phys. Rev. Lett.}\ }\textbf {\bibinfo {volume} {107}},\ \bibinfo
  {pages} {240501} (\bibinfo {year} {2011})}\BibitemShut {NoStop}%
\bibitem [{\citenamefont {Morse}\ \emph {et~al.}(1999)\citenamefont {Morse},
  \citenamefont {Hamilton}, \citenamefont {Johnson}, \citenamefont {Mauceli},\
  and\ \citenamefont {McHugh}}]{lsu}%
  \BibitemOpen
  \bibfield  {author} {\bibinfo {author} {\bibfnamefont {A.}~\bibnamefont
  {Morse}}, \bibinfo {author} {\bibfnamefont {W.~O.}\ \bibnamefont {Hamilton}},
  \bibinfo {author} {\bibfnamefont {W.~W.}\ \bibnamefont {Johnson}}, \bibinfo
  {author} {\bibfnamefont {E.}~\bibnamefont {Mauceli}}, \ and\ \bibinfo
  {author} {\bibfnamefont {M.~P.}\ \bibnamefont {McHugh}},\ }\href@noop {}
  {\bibfield  {journal} {\bibinfo  {journal} {Phys. Rev. D}\ }\textbf {\bibinfo
  {volume} {59}},\ \bibinfo {pages} {062002} (\bibinfo {year}
  {1999})}\BibitemShut {NoStop}%
\bibitem [{\citenamefont {Reagor}\ \emph {et~al.}(2013)\citenamefont {Reagor},
  \citenamefont {Paik}, \citenamefont {Catelani}, \citenamefont {Sun},
  \citenamefont {Axline}, \citenamefont {Holland}, \citenamefont {Pop},
  \citenamefont {Masluk}, \citenamefont {Brecht}, \citenamefont {Frunzio},
  \citenamefont {Devoret}, \citenamefont {Glazman},\ and\ \citenamefont
  {Schoelkopf}}]{reagor}%
  \BibitemOpen
  \bibfield  {author} {\bibinfo {author} {\bibfnamefont {M.}~\bibnamefont
  {Reagor}}, \bibinfo {author} {\bibfnamefont {H.}~\bibnamefont {Paik}},
  \bibinfo {author} {\bibfnamefont {G.}~\bibnamefont {Catelani}}, \bibinfo
  {author} {\bibfnamefont {L.}~\bibnamefont {Sun}}, \bibinfo {author}
  {\bibfnamefont {C.}~\bibnamefont {Axline}}, \bibinfo {author} {\bibfnamefont
  {E.}~\bibnamefont {Holland}}, \bibinfo {author} {\bibfnamefont {I.~M.}\
  \bibnamefont {Pop}}, \bibinfo {author} {\bibfnamefont {N.~A.}\ \bibnamefont
  {Masluk}}, \bibinfo {author} {\bibfnamefont {T.}~\bibnamefont {Brecht}},
  \bibinfo {author} {\bibfnamefont {L.}~\bibnamefont {Frunzio}}, \bibinfo
  {author} {\bibfnamefont {M.~H.}\ \bibnamefont {Devoret}}, \bibinfo {author}
  {\bibfnamefont {L.}~\bibnamefont {Glazman}}, \ and\ \bibinfo {author}
  {\bibfnamefont {R.~J.}\ \bibnamefont {Schoelkopf}},\ }\href@noop {}
  {\bibfield  {journal} {\bibinfo  {journal} {Applied Physics Letters}\
  }\textbf {\bibinfo {volume} {102}} (\bibinfo {year} {2013})}\BibitemShut
  {NoStop}%
\bibitem [{\citenamefont {Miroshnichenko}, \citenamefont {Flach},\ and\
  \citenamefont {Kivshar}(2010)}]{RevModPhys.82.2257}%
  \BibitemOpen
  \bibfield  {author} {\bibinfo {author} {\bibfnamefont {A.~E.}\ \bibnamefont
  {Miroshnichenko}}, \bibinfo {author} {\bibfnamefont {S.}~\bibnamefont
  {Flach}}, \ and\ \bibinfo {author} {\bibfnamefont {Y.~S.}\ \bibnamefont
  {Kivshar}},\ }\href@noop {} {\bibfield  {journal} {\bibinfo  {journal}
  {Reviews of Modern Physics}\ }\textbf {\bibinfo {volume} {82}},\ \bibinfo
  {pages} {2257} (\bibinfo {year} {2010})}\BibitemShut {NoStop}%
\bibitem [{\citenamefont {Lepetit}\ \emph {et~al.}(2010)\citenamefont
  {Lepetit}, \citenamefont {Akmansoy}, \citenamefont {Ganne},\ and\
  \citenamefont {Lourtioz}}]{PhysRevB.82.195307}%
  \BibitemOpen
  \bibfield  {author} {\bibinfo {author} {\bibfnamefont {T.}~\bibnamefont
  {Lepetit}}, \bibinfo {author} {\bibfnamefont {E.}~\bibnamefont {Akmansoy}},
  \bibinfo {author} {\bibfnamefont {J.-P.}\ \bibnamefont {Ganne}}, \ and\
  \bibinfo {author} {\bibfnamefont {J.-M.}\ \bibnamefont {Lourtioz}},\
  }\href@noop {} {\bibfield  {journal} {\bibinfo  {journal} {Physical Review
  B}\ }\textbf {\bibinfo {volume} {82}},\ \bibinfo {pages} {195307} (\bibinfo
  {year} {2010})}\BibitemShut {NoStop}%
\bibitem [{\citenamefont {Frigola}\ \emph {et~al.}(2015)\citenamefont
  {Frigola}, \citenamefont {Agustsson}, \citenamefont {Faillace}, \citenamefont
  {A.}, \citenamefont {Ciovati}, \citenamefont {Clemens}, \citenamefont
  {Dhakal}, \citenamefont {Marhauser}, \citenamefont {Rimmer}, \citenamefont
  {Spradlin},\ and\ \citenamefont {S.}}]{niobium}%
  \BibitemOpen
  \bibfield  {author} {\bibinfo {author} {\bibfnamefont {P.}~\bibnamefont
  {Frigola}}, \bibinfo {author} {\bibfnamefont {R.}~\bibnamefont {Agustsson}},
  \bibinfo {author} {\bibfnamefont {L.}~\bibnamefont {Faillace}}, \bibinfo
  {author} {\bibfnamefont {M.}~\bibnamefont {A.}}, \bibinfo {author}
  {\bibfnamefont {G.}~\bibnamefont {Ciovati}}, \bibinfo {author} {\bibfnamefont
  {W.}~\bibnamefont {Clemens}}, \bibinfo {author} {\bibfnamefont
  {P.}~\bibnamefont {Dhakal}}, \bibinfo {author} {\bibfnamefont
  {F.}~\bibnamefont {Marhauser}}, \bibinfo {author} {\bibfnamefont
  {R.}~\bibnamefont {Rimmer}}, \bibinfo {author} {\bibfnamefont
  {J.}~\bibnamefont {Spradlin}}, \ and\ \bibinfo {author} {\bibfnamefont
  {W.}~\bibnamefont {S.}},\ }in\ \href@noop {} {\emph {\bibinfo {booktitle}
  {Proceedings of SRF2015, Whistler, BC, Canada}}}\ (\bibinfo {year}
  {2015})\BibitemShut {NoStop}%
\end{thebibliography}%

\end{document}